\documentclass[12pt]{article}
\usepackage{epsfig}
\usepackage{multirow}

\def\beq{\begin{equation}}
\def\eeq#1{\label{#1}\end{equation}}
\def\eeqn{\end{equation}}

\def\beqa{\begin{eqnarray}}
\def\eeqa#1{\label{#1}\end{eqnarray}}
\def\eeqan{\end{eqnarray}}

\let\bar=\overbar

\def\Dslash{\not{\hbox{\kern-4pt $D$}}}
\def\dslash{\not{\hbox{\kern-2pt $\del$}}}

\def\msb{{\bar{\ssstyle M \kern -1pt S}}}

\def\Title#1{\begin{center} {\Large {\bf #1} } \end{center}}

\begin{document}

\Title{SuperB and BelleII prospects on direct $CP$ violation measurements}

\bigskip\bigskip

%+\addtocontents{toc}{{\it D. Reggiano}}
%+\label{ReggianoStart}

\begin{raggedright}  

{\it Yoshiyuki Onuki\index{Onuki, Y.}\\
Department of Physics\\
University of Tokyo\\
7-3-1 Hongo, Bunkyo-ku, 113-0033 Tokyo, JAPAN}
\bigskip\bigskip
\end{raggedright}

\section{Introduction}

\begin{figure}[htbp]
 \begin{center}
 \includegraphics*[width=0.7\textwidth]{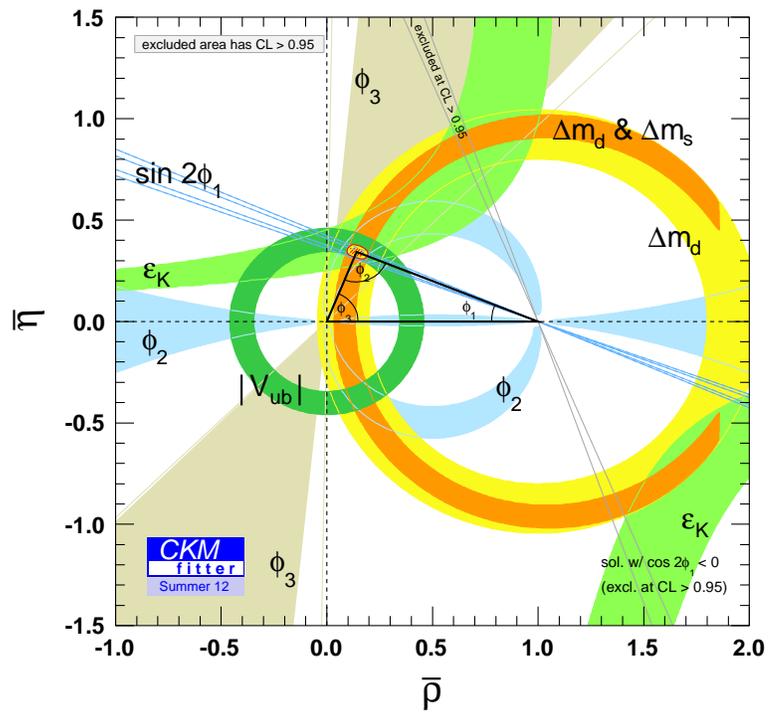}
 \caption{\sl
 The unitarity triangle.
 \label{fig:triangle}
 }
 \end{center}
\end{figure}

The $CP$ violation observed in quark sector is 
explained by a irreducible complex phase in the Cabibbo-Kobayashi-Maskawa(CKM) 
matrix\cite{ref:ckm} in the Standard Model. 
One of the unitarity constraints of the CKM matrix is given by the equation
$V_{ud}V_{ub}^{*}+V_{cd}V_{cb}^{*}+V_{td}V_{tb}^{*}=0$ 
which is represented by a triagnle in the complex plane.
The phase can be determined from measurements of the three angles and sides
of the triangle.
The angles are called as
$\alpha/\phi_{2} = Arg[-(V_{td}V_{Vtb}^{*})/(V_{ud}V_{ub}^{*})]$, 
$\beta/\phi_{1} = Arg[-(V_{cd}V_{cb}^{*})/(V_{td}V_{tb}^{*})]$ and 
$\gamma/\phi{3} = Arg[-(V_{ud}V_{ub}^{*})/(V_{cd}V_{cb}^{*})]$. 
The latest results of the measurements of sides and angles are shown in the Fig.\ref{fig:triangle}.

%\begin{figure}[htb]
%\centering
%\includegraphics[height=1.5in]{diagram}
%\caption{(a)Cabibbo favored decay, (b)Doubly Cabibbo suppressed decay in $B\rightarrow D^{(*)}\pi$}
%\label{fig:diagram}
%\end{figure}

The angle $\gamma/\phi_{3}$ is the least well determined among all angles.
%In the time-dependent $\gamma/\phi_{3}$ measurement of $D^{(*)}\pi$, 
%at first $CP$ violation phase $2\beta/\phi_{1}$ is appeared in the $B\bar{B}$ 
%mixing process and $\gamma/\phi_{3}$ arised from followed two decay paths: 
%Cabibbo favored decay (CFD) and doubly Cabibbo suppressed decay (DCSD) 
%into a final state.
%Improvement of the precision is eagerly awaited.
The measurement of $\gamma$/$\phi_{3}$ had been proposed to use the process 
$B\rightarrow D^{(*)}K^{(*)}$ involved with interference with
$b\rightarrow u$ and $b\rightarrow c$ 
quark transition in the discussion of direct $CP$ violation\cite{ref:sanda}. 
The measurement of $\gamma/\phi_{3}$ is performed in a theoretically cleanly 
way since only the tree-dominated decays are involved.
Some methods to extract $\gamma$/$\phi_{3}$ had been suggested so far:
GLW\cite{ref:GLW}, ADS\cite{ref:ADS},  Dalitz\cite{ref:ggsz, ref:bondar}
analyses.
The GLW analysis uses $D^{0}$ and $\bar{D}^{0}$ decay into $CP$ eigenstates
such as $K^{+}K^{-}$ or $K_{S}\pi^{0}$, etc.
The observables of double ratio and asymmetry are defined as below:
\begin{eqnarray}
\displaystyle
\begin{array}{rll}
R_{CP\pm} &\equiv & \displaystyle 2\frac{{\cal B}(B^{-}\rightarrow D_{CP\pm}K^-)  + {\cal B}(B^{-}\rightarrow D_{CP\pm}K^{-})}{{\cal B}(B^{-}\rightarrow D^0K^{-})  + {\cal B}(B^{-}\rightarrow \bar{D}^0K^{-})}\\
& = & 1 + r_{B}^{2} \pm 2r_{B}\cos\delta_{B}\cos\phi_{3}
\end{array}
\end{eqnarray}
\begin{eqnarray}
\displaystyle
\begin{array}{rll}
A_{CP\pm} &\equiv &\displaystyle\frac{{\cal B}(B^{-}\rightarrow D_{CP\pm}K^-) - {\cal B}(B^{+}\rightarrow D_{CP\pm}K^+)}{{\cal B}(B^{-}\rightarrow D_{CP\pm}K^-) + {\cal B}(B^{+}\rightarrow D_{CP\pm}K^+)}\\
& = & \pm2r_{B}\sin\delta_{B}\sin\phi_{3}/R_{CP\pm}
\end{array}
\end{eqnarray}
where the $D_{CP\pm}$ is 
the $D$ meson reconstructed in the $CP$-even(+) or $CP$-odd(-) final state, 
$r_{B}$ is the ratio of amplitudes between $B^{-}\rightarrow \bar{D}^{0}K^{-}$
and
$B^{-}\rightarrow D^{0}K^{-}$ defined as
 $r_{B} \equiv |A(B^{-}\rightarrow \bar{D}^{0}K^{-})/|A(B^{-}\rightarrow D^{0}K^{-})|$, 
and $\delta_{B}$ is the difference of strong phase for these amplitudes.

%The ADS analysis uses 
%Cabibbo-favored $\bar{D}^{0}$ decay(CFD) and 
%doubly Cabibbo-suppressed $D^{0}$ decay(DCSD) to adjust the interfering
%amplitudes have comparable magnitudes through the 
The ADS analysis uses 
$B^{-}\rightarrow D^{(*)}K^{(*)-}$ decays followed by the
Cabibbo-favored(CFD) and doubly Cabibbo-suppressed $D^{0}$ decays(DCSD), 
where the interfering amplitudes have comparable magnitude.
The CFD(DCFD) decays of the $D$ meson that can be used for ADS 
are $D^{0}\rightarrow K^{-}\pi^{+}$, $K^{-}\pi^{+}\pi^{0}$
($D^{0}\rightarrow K^{+}\pi^{-}$, $K^{+}\pi^{-}\pi^{0}$), etc.
The observables, double ratio and asymmetry, are defined as below.
\begin{eqnarray}
\displaystyle
\begin{array}{rll}
R_{ADS} &\equiv & \displaystyle\frac{{\cal B}(B^{-}\rightarrow [f]_{D}K^-)  + {\cal B}(B^{+}\rightarrow [\bar{f}]_{D}K^{+})}{{\cal B}(B^{-}\rightarrow [\bar{f}]_{D}K^{-})  + {\cal B}(B^{+}\rightarrow [f]_{D}K^{+})}\\
& = & r_{B}^{2} + r_{D}^{2} + 2r_{B}r_{D}\cos(\delta_{B}+\delta_{D})\cos\phi_{3}
\end{array}
\end{eqnarray}

\begin{eqnarray}
\displaystyle
\begin{array}{rll}
A_{ADS} &\equiv & \displaystyle\frac{{\cal B}(B^{-}\rightarrow [f]_{D}K^-)  - {\cal B}(B^{+}\rightarrow [\bar{f}]_{D}K^{+})}{{\cal B}(B^{-}\rightarrow [\bar{f}]_{D}K^{-})  + {\cal B}(B^{+}\rightarrow [f]_{D}K^{+})}\\
& = & 2r_{B}r_{D}\sin(\delta_{B}+\delta_{D})\sin\phi_{3}/R_{ADS}
\end{array}
\end{eqnarray}
where $r_{D}=|A(D^{0}\rightarrow f)/A(\bar{D}^{0}\rightarrow f)|$ and
$\delta_{D}$ is strong phase difference between 
$\bar{D}^{0}\rightarrow f$ and $D^{0}\rightarrow f$.

The Dalitz analysis with $D$ meson decay into 
the three-body decay $K_{S}h^{+}h^{-}$ to extract the angle 
$\gamma$/$\phi_{3}$, where $h^{\pm}$ represents charged light hadrons 
such as pion and kaon.
The model-dependent Datliz analysis uses the isobar model~\cite{ref:isobar}
which assume that the three-body decay of the $D$ meson proceeds through 
the intermediate two-body resonances. 
The total amplitude over the Dalitz plot can be represented 
as the sum of two amplitudes for $D^{0}$ and $\bar{D}^{0}$ decays 
into the same final state $K_{S}h^{+}h^{-}$ 
as below.
\begin{eqnarray}
\displaystyle
\begin{array}{r}
f_{B^{+}} = f_{D}(m_{+}^{2}, m_{-}^{2}) + r_{B}e^{\pm i\phi_{3}+i\delta_{B}}f_{D}(m_{-}^{2},m_{+}^{2})\\
\end{array}
\end{eqnarray}
where $m_{+}^{2} = m_{K_{S}h^{+}}^{2}$, $m_{-}^{2} = m_{K_{S}h^{-}}^{2}$.
The $f_{D}(m_{+}^{2},m_{-}^{2})$ consists of the sum of
intermediates two-body amplitudes and a single non-resonant 
amplitude as follows.
\begin{eqnarray}
\displaystyle
\begin{array}{r}
f_{D}(m_{+}^{2},m_{-}^{2}) = \displaystyle\sum^{N}_{j=1}a_{j}e^{i\xi_{j}}{\cal A}_{j}(m_{+}^{2},m_{-}^{2})+a_{NR}e^{i\xi_{NR}} 
\end{array}
\end{eqnarray}
Where 
$a_{j}$ and $\xi_{j}$ are the amplitude and phase of the matrix element,
${\cal A}_{j}$ is the matrix element of the $j$-th resonance, and $a_{NR}$ and
$\xi_{NR}$ are the amplitude and phase of the non-resonant component.
The $r_{B}e^{\pm i\phi_{3}+i\delta_{B}}$ can be converted to the 
Cartesian parameters $x_{\pm}=r_{\pm}\cos(\pm \phi_{3}+\delta)$ and 
$y_{\pm}=r_{\pm}\sin(\pm\phi_{3}+\delta)$. The $x_{\pm}$ and $y_{\pm}$ are
actual fitted parameters.

The precision of $\gamma$/$\phi_3$ measurement has progressed 
as the data accumulated at $B$-factories 
and new efficient physics methods and analysis techniques were developed. 
Although current statistics of $e^+e^-$ colliders is over the 1.2 billion 
$B\bar{B}$ pairs, it is insufficient for reliable $\gamma$/$\phi_{3}$.
The SuperB\cite{ref:SuperB} and BelleII\cite{ref:BelleII} projects are planned to accumulate the 50-75 times larger than at the $B$-factories with improved detectors in the next decade.

\section{$\gamma/\phi_{3}$ determination}

The current most precise determination of $\gamma/\phi_{3}$ have been performed
with Dalitz method. The $\gamma/\phi_{3}$ results of an model-dependent unbinned Dalitz method by Belle and BaBar are $\gamma/\phi_{3}=(78.4^{+10.8}_{-11.6} \pm 3.6 \pm 8.9)^{\circ}$ \cite{ref:belle_model-dep2}, $(68 \pm 14 \pm 4 \pm 3)^{\circ}$\cite{ref:babar_model-dep3} with modulo $180^\circ$, respectively, 
where the 3rd error is the model uncertainty. 
The model-dependent measurement is likely to become dominated by the model uncertainty in the Super $B$-factories era. 
The new technique using model-independent binned Dalitz method\cite{ref:bonder_model-indep1,ref:bonder_model-indep2} is reported by Belle\cite{ref:belle_model-indep} is supposed to eliminate this uncertainty.
The model-independent Dalitz plot is divided into 2$N$ bins symmetrically under
the exchange $m^{2}_{-} \leftrightarrow m^2_{+}$. 
The bin index $i$ ranges from $-N$ to $N$ excluding 0. 
The expected number of events in bin $i$ of the Dalitz
plot of the $D$ meson from $B^{\pm}\rightarrow DK^{\pm}$ is
\begin{eqnarray}
\begin{array}{l}
N^{\pm}_{i} = h_{B}[K_{\pm i} + r_{B}^2 K_{\mp i}+2\sqrt{K_{i}K_{-i}}(x_{\pm}c_{i}\pm y_{\pm}s_{i})]
\end{array}
\label{eqn:model-independent_Dalitz}
\end{eqnarray}
where $h_{B}$ is a normalization constant and $K_{i}$ is the number of events
in the $i$th bin of the $K_{S}^{0}\pi^{+}\pi^{-}$ Dalitz plot of the $D$ meson
in a flavor eigenstate. $x_{\pm}$ and $y_{\pm}$ are the same parameters as the
ones used in the model-dependent Dalitz analyses.

The Belle reported the first $\gamma/\phi_{3}$ measurement with the model-dependent Dalitz method,
$\gamma/\phi_{3} = (77.3^{+15.1}_{-14.9}\pm 4.1 \pm 4.3(c_{i},s_{i}))^{\circ}$. 
Here the 3rd error is the uncertainty of the strong phase determination in the Dalitz plane studied by CLEOc experiment based on 818$pb^{-1}$ at $\Upsilon(3770)$\cite{ref:cleo_cisi}.
Now, the BESIII\cite{ref:BESIII_detector} experiment had accumulated the 2.9 $fb^{-1}$. The uncertainty is expected to be less than 1 degree in the near future.

\begin{figure}[htbp]
 \begin{center}
 \includegraphics*[width=0.7\textwidth]{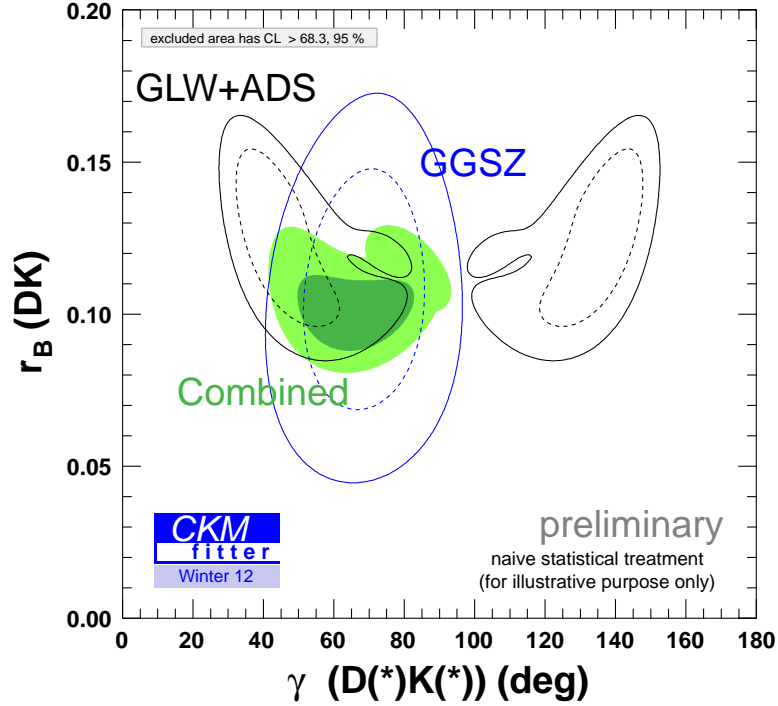}
 \caption{\sl 
   The correlation between the $\gamma/\phi_{3}$ and the ratio of interfering amplitudes $r_{B}$ of the decay $B\rightarrow DK$ from world average $D^{(*)}K^{(*)}$ decays (GLW+ADS) and Dalitz analyses.
 \label{fig:rBvsgamma_DK}
 }
 \end{center}
\end{figure}

The GLW and ADS combined measurements have also comparable constraints and model-independent on the $\gamma/\phi_{3}$ determination in Fig.\ref{fig:rBvsgamma_DK}.
These measurements have a important role to determine the $\gamma/\phi_{3}$ in the SuperB and BelleII era. The expected precisions are shown in the Table.\ref{tab:blood}.

\begin{table}[htbp]
\begin{center}
\begin{tabular}{lccc}  
\hline
\hline
Observable &  $B$ Factories(2$ab^{-1}$) & SuperB(75$ab^{-1}$) & BelleII(50$ab^{-1}$)  \\ 
\hline
$\gamma/\phi_{3}$($B\rightarrow DK, GLW$) &  $\sim$ 15$^\circ$ & 2.5$^\circ$ & \multirow{2}{*}{\} 5$^\circ$}  \\
$\gamma/\phi_{3}$($B\rightarrow DK, ADS$)  &  $\sim$12$^\circ$ & 2.0$^\circ$ &  \\
$\gamma/\phi_{3}$($B\rightarrow DK, Dalitz$)  &  $\sim$9$^\circ$ & 1.5$^\circ$ &  2$^\circ$ \\
$\gamma/\phi_{3}$($B\rightarrow DK, combined$)  &  $\sim$6$^\circ$ & 1-2$^\circ$ & 1.5$^\circ$ \\
\hline
\end{tabular}
\caption{The expected precision of $\gamma/\phi_3$ determination at SuperB and BelleII. Both the statistical and systematic errors are assumed to scale with the integrated luminosity.}
\label{tab:blood}
\end{center}
\end{table}

\section{Direct $CP$ violation in charmeless hadronic decay}

\begin{figure}[htbp]
 \begin{center}
 \includegraphics*[width=0.7\textwidth]{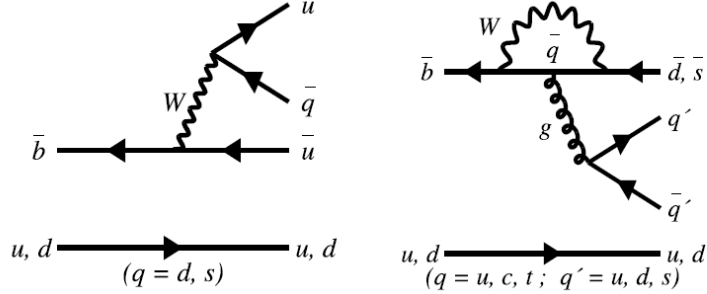}
 \caption{\sl 
   Tree diagram and penguin diagram in $B\rightarrow hh$ decay.
 \label{fig:btohh_diagram}
 }
 \end{center}
\end{figure}

The direct CP asymmetries has been observed in two-body decays such as  
$B^{0}\rightarrow \pi\pi$ and 
$B^{0}\rightarrow K\pi$ decays.
The charmless 2-body $B$ meson decays could receive contribution from 
processes beyond the standard model. 
For example, the $B\rightarrow K\pi$ proceeds through the suppressed tree diagram and loop penguin diagram of similar magnitude(Fig.\ref{fig:btohh_diagram}). 
The interference of the two diagrams cause a direct $CP$ asymmetry of $A_{CP}^{f} = [\Gamma(\bar{B}\rightarrow\bar{f})-\Gamma(B\rightarrow f)]/[\Gamma(\bar{B}\rightarrow\bar{f})+\Gamma(B\rightarrow f)]$ . 

\begin{figure}[htbp]
 \begin{center}
 \includegraphics*[width=0.7\textwidth]{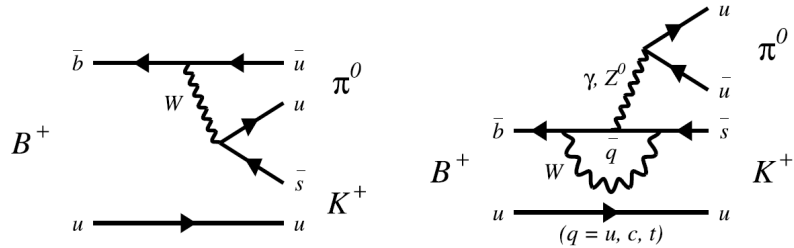}
 \caption{\sl 
   Color suppressed diagram and electroweak penguin diagram in $B^{+}\rightarrow K^{+}\pi^{0}$ decay.
 \label{fig:btohh_diagram2}
 }
 \end{center}
\end{figure}

The processes involved in the decays of neutral and charged $B$ decays to $K\pi$
are expected to be the same.
Additional diagrams which can contribute to $B^{+}$ decays shown in Fig.\ref{fig:btohh_diagram2} are expected to be much smaller than the contributions in Fig.\ref{fig:btohh_diagram}, thus, the asymmetries $A_{CP}^{K^{+}\pi^{0}}$ in $B^{\pm}\pi^{0}$ decays and $A_{CP}^{K^{+}\pi^{-}}$ in $B^{0}(\bar{B}^{0})\rightarrow K^{\pm}\pi^{\mp}$ decays are expected to be the same. 
The recent averages of $A_{CP}^{K^{+}\pi^{0}}$ and $A_{CP}^{K^{+}\pi^{-}}$ 
by HFAG\cite{ref:HFAG} show significant (5$\sigma$) deviation of $\Delta A_{K\pi}=A_{CP}^{K^{+}\pi^{0}}-A_{CP}^{K^{+}\pi^{-}}$ from $0$.
This is known as $\Delta A^{K\pi}$ puzzle.
A sum rule relation\cite{ref:sum_rule} in Equation.\ref{eqn:sum_rule} proposed to test the puzzle with various measured observables in $K\pi$ decays.

\begin{eqnarray}
\begin{array}{l}
A_{CP}^{K^{+}\pi^{-}}
+A_{CP}^{K^{0}\pi^{+}}\frac{{\cal B}(B^{+}\rightarrow K^{0}\pi^{+})\tau_{B^{0}}}
{{\cal B}(B^{0}\rightarrow K^{+}\pi^{-})\tau_{B^{+}}}
=A_{CP}^{K^{+}\pi^{0}}\frac{2{\cal B}(B^{+}\rightarrow K^{+}\pi^{0})\tau_{B^{0}}}{{\cal B}(B^{0}\rightarrow K^{+}\pi^{-})\tau_{B^{+}}}+A_{CP}^{K^{0}\pi^{0}}\frac{2{\cal B}(B^{0}\rightarrow K^{0}\pi^{0})}{{\cal B}(B^{0}\rightarrow K^{+}\pi^{-})}\\
\end{array}
\label{eqn:sum_rule}
\end{eqnarray}

where ${\cal B}(B\rightarrow f)$ denotes the corresponding branching fraction and $\tau_{B^{0}(B^{+})}$ life time of neutral and charged $B$ mesons.

\begin{figure}[htbp]
 \begin{center}
 \includegraphics*[width=0.8\textwidth]{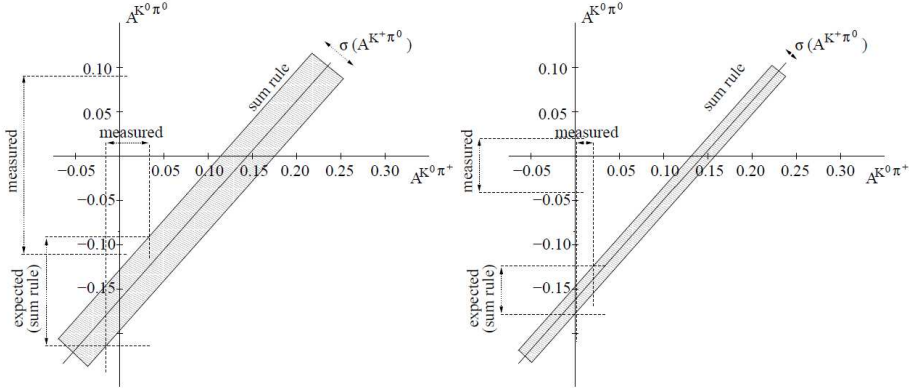}
 \caption{\sl 
   Current world-average constraints on $A_{CP}^{K^{0}\pi^{0}}$ vs $A_{CP}^{K^{0}\pi^{+}}$\cite{ref:HFAG}(left). Expected constraints with the same central values and scaled for the integrated luminosity of $L$ = 50 $ab^{-1}$ at the BelleII. 
Contribution of systematic error in the figure adapts  
the current systematic error without any scaling conservatively.
 \label{fig:AcpSumRule_BelleII}
 }
 \end{center}
\end{figure}

Using the current world average values for the corresponding observables\cite{ref:HFAG}, the sum rule can be represented as a dependence of the least precise asymmetry $A_{CP}^{K^{0}\pi^{0}}$ on the $A_{CP}^{K^{0}\pi^{+}}$ as shown in Fig.\ref{fig:AcpSumRule_BelleII}. A violation of the sum rule would indicate new physics in $b\rightarrow \bar{q}q$ transition.

\section{Conclusion}

Current most precise determination of $\gamma/\phi_{3}$ 
is brought by the Dalitz analyses.
Both the model-independent and improved model-dependent analysis pushed down 
the systematic limitation and open up the possibilities of much higher 
precision determination at super $B$ factories in near future.
Furthermore, the combined ADS and GLW results have a competitive determination with the Dalitz analysis. 
Since the measurement of $\gamma/\phi_{3}$ is obtained theoretically cleanly from the tree-dominated decays, the precise measurement will still play a important role for the test of unitarity triangle in the super $B$ factories era.

The direct $CP$ violation in the charmless hadronicis decay is suitable  
place to explore the new physics phenomena.
A violation of the sum rule of $B\rightarrow K\pi$ would indicate new physics in $b\rightarrow \bar{q}q$ transition. 
The SuperB and BelleII also have good potential to search for the existence of new physics in this mode.

\end{document}